\documentclass[aps,pra,10pt]{revtex4-2}
\usepackage{graphicx} 
\usepackage{physics}
\usepackage{amsmath}
\usepackage{setspace}
\usepackage{anysize} 
\usepackage{xcolor}
\usepackage{braket}
\usepackage{amsfonts} 
\usepackage{microtype}
\usepackage[normalem]{ulem}
\usepackage{mathtools}
\usepackage{bm}

\usepackage[colorlinks=true, linkcolor=black, citecolor=black, urlcolor=blue]{hyperref}
\graphicspath{ {./figures/} }

\begin{document}

\preprint{APS/123-QED}

\title{Exact versus tight-binding models in longitudinally modulated \texorpdfstring{$\mathcal{PT}$}{PT}-symmetric coupled waveguides}

\author{Alonso Contreras-Astorga}
\email{alonso.contreras@cinvestav.mx}
\affiliation{SECIHTI - Physics Department, Cinvestav, P.O. Box 14-740, 07000, Mexico City, Mexico.}

\author{José Israel Galindo-Rodríguez}
\email{jose.galindo@cinvestav.mx}
\affiliation{Physics Department, Cinvestav, P.O. Box 14-740, 07000, Mexico City, Mexico.}

\begin{abstract}
The tight-binding (TB) model is a widely adopted approximation scheme for describing light propagation in waveguide arrays. Despite its success, its validity in $\mathcal{PT}$-symmetric systems characterized by strong longitudinal modulation has not been rigorously benchmarked against exact analytical solutions. In this work, we address this gap by performing a comparative analysis between exact continuous solutions derived from $z$-dependent supersymmetric (SUSY) transformations and their corresponding discrete TB approximations. To achieve this, we develop a theoretical model for two $\mathcal{PT}$-symmetric coupled waveguides subject to longitudinal modulation. We then evaluate the performance of the TB framework against the exact SUSY benchmark. Our results delineate the specific validity range of the TB approximation, demonstrating its proficiency in reproducing spatial intensity distributions. However, we also identify its limitations in accurately capturing the complex oscillatory phase dynamics inherent to this non-Hermitian evolution.
\end{abstract}

\maketitle

\section{Introduction} 

The mathematical equivalence between paraxial light propagation and the Schrödinger equation has established optics as an excellent platform for the simulation of synthetic quantum systems \cite{glogeFormalQuantumTheory1969,laxMaxwellParaxialWave1975,cruzycruz2017GroupApproachParaxial}. Within this formal correspondence, the longitudinal coordinate $ z $ acts as an effective temporal parameter, while the transverse refractive index profile mimics a potential energy distribution. This space-time mapping is particularly powerful for studying non-Hermitian Hamiltonians obeying Parity-Time ($ \mathcal{PT}$) symmetry. In such systems, a balanced spatial distribution of gain and loss—implemented via complex refractive index landscapes \cite{makrisPTSymmetricOptical2010,miriSupersymmetricOpticalStructures2013}—replaces the conventional requirement of Hermiticity while still preserving entirely real energy spectra \cite{bender1998RealSpectraNonHermitian,znojilShouldPTSymmetric2002,mostafazadehPseudoHermiticityPTSymmetry2002,mostafazadehExactPTSymmetry2003,ruterObservationParityTime2010,mannheimAppropriateInnerProduct2018}.

Integrated optical waveguide arrays have been widely explored as a platform for the quantum simulation of physical systems \cite{longhiQuantumopticalAnalogiesUsing2009}. While early research primarily addressed static configurations \cite{heinrich2014SupersymmetricModeConverters,correa2015SymmetricInvisibleDefects,queralto2017ModedivisionDemultiplexingUsing}, recent focus has shifted toward longitudinal modulation — where the waveguide parameters vary along the propagation axis $z$. This $ z $-dependent evolution facilitates an additional degree of freedom to steer light, enabling phenomena such as dynamic localization and self-collimation \cite{longhiSelfcollimationSelfimagingEffects2008}, hyper-ballistic transport via periodic modulation \cite{dellavalleSpectralTransportProperties2013}, Floquet engineering of light localization \cite{maFloquetEngineeringLocalized2018}, and the design of advanced periodic structures \cite{raja2020PhaseshiftedPTSymmetric,liuPerfectExcitationTopological2024}. However, when non-Hermitian $ \mathcal{PT} $ symmetry is combined with such longitudinal variations, the analytical description of the system becomes significantly more intricate. The interplay between the balanced gain-loss profile and the $z$-dependent modulation necessitates robust theoretical frameworks capable of handling non-Hermitian Hamiltonians. In particular, it remains unclear to what extent standard approximation methods can accurately capture the combined effects of non-Hermiticity and longitudinal modulation.

Two distinct theoretical frameworks are typically employed to analyze these complex systems. On the one hand, Supersymmetry (SUSY) or Darboux transformations offer a rigorous mathematical approach to construct exactly solvable potentials with tailored spectral properties \cite{matveev1991DarbouxTransformationsSolitons,cooper1995SupersymmetryQuantumMechanicsa,junker1996SupersymmetricMethodsQuantum,fernandezc.HigherorderSupersymmetricQuantum2004,miriSupersymmetrygeneratedComplexOptical2013}. While originally developed for stationary regimes, these transformations have been successfully extended to the $ z $-dependent Schrödinger equation, providing exact analytical solutions even for complex non-Hermitian Hamiltonians \cite{bagrovDarbouxTransformationNonsteady1995,bagrovSupersymmetryNonstationarySchrodinger1996,contreras-astorgaTimeDependentAnharmonicOscillator2017,cenTimedependentDarbouxSupersymmetric2019,contreras-astorgaPhotonicSystemsTwodimensional2019,contreras-astorgaMultimodeTwoDimensionalPTSymmetric2020,cruzycruzCoherentStatesExactly2020}. On the other hand, the Tight-Binding (TB) model remains one of the standard approximation schemes for describing light propagation in waveguide arrays \cite{dittrichQuantumTransportDissipation1998,amiriTightBindingAnalysisCoupled2006}. By assuming that the evolution is governed by discrete hopping between localized modes, the TB framework provides intuitive physical insights and computational efficiency \cite{munoz-vegaExactlySolubleModel2014,principeSupersymmetryInspiredNonHermitianOptical2015,kittelIntroductionSolidState2018,chenTightbindingModelOptical2021}, reducing the continuous system to a set of coupled differential equations.

Despite the widespread adoption of the TB model, its validity in $ \mathcal{PT} $ symmetric systems characterized by strong longitudinal modulation has not been rigorously benchmarked against exact analytical solutions. In this paper, we address the comparative analysis between exact solutions derived from $ z $-dependent Darboux transformations and the corresponding TB approximations. We develop a theoretical model for two $ \mathcal{PT}$-symmetric coupled waveguides with longitudinal modulation and evaluate the performance of the TB model against the exact SUSY benchmark. Our results delineate the validity range of the TB approximation, demonstrating its proficiency in reproducing spatial intensity distributions while identifying its limitations in capturing the complex oscillatory phase dynamics inherent to non-Hermitian evolution.

\section{Notes on supersymmetric quantum mechanics} \label{TDDT}

This section outlines the formalism of supersymmetric quantum mechanics, focusing on both time-dependent and time-independent frameworks. First, we introduce first-order transformations to generate new exactly solvable potentials, discussing the specific constraints required to preserve Hermiticity and $\mathcal{PT}$ symmetry. Subsequently, we extend this formalism to second-order transformations, demonstrating how regular, non-singular physical potentials can be constructed even when the intermediate states exhibit singularities.

\subsection{Time-dependent supersymmetric transformations}

Consider the time-dependent Schrödinger equation
\begin{equation} \label{TDDT-S0Equation}
    S_0 \psi_0 (x,z) = 0, 
\end{equation}
where $S_0 = i \partial_z + \partial_x^2 - V_0(x,z)$ is the Schrödinger operator, and the propagation variable $z$ plays the role of time. We assume that the initial potential $V_0$ is exactly solvable and that its corresponding solutions $\psi_0$ are known. We can employ time-dependent Darboux transformations to construct a new exactly solvable Schrödinger operator $S_1$ characterized by a potential $V_1(x, z)$ and solutions $\psi_1(x, z)$. To achieve this, we introduce the intertwining relation
\begin{equation} \label{TDDT-S0S1IntertwineRel}
    S_1 L_1 = L_1 S_0,
\end{equation}
which guarantees that $\psi_1 = L_1 \psi_0$ is a solution to $S_1 \psi_1 (x, z) = 0$. The ansatz for the intertwining operator is chosen as
\begin{equation} \label{TDDT-L1Operator}
    L_1 (x,z) = \ell_1 (z) \left ( \partial_x - \frac{ \partial_x u_1 (x,z) }{ u_1 (x,z) } \right ).
\end{equation}
Substituting Eq.~\eqref{TDDT-L1Operator} into Eq.~\eqref{TDDT-S0S1IntertwineRel} and solving for $V_1$, we obtain
\begin{equation} \label{TDDT-V1}
    V_1 (x,z) =  V_0 (x,z) - 2 \partial_x^2 \ln u_1 (x,z) +  i \partial_z \ln \ell_1 (z),
\end{equation} 
where $u_1(x,z)$ is a solution to Eq.~\eqref{TDDT-S0Equation}, commonly referred to as the transformation function, which must be nodeless to avoid singularities in $V_1$, and $\ell_1(z)$ is an arbitrary, non-vanishing function of $z$.

In Ref.~\cite{bagrovDarbouxTransformationNonsteady1995}, it was required that both $V_0$ and $V_1$ be real to preserve the Hermiticity of the operator $S_1$. This requirement leads to the following condition for the functions $\ell_1$ and $u_1$:
\begin{equation} \label{TDDT-HermiticCondition}
    2 \partial_x^2 \ln \frac{u_1(x,z)}{u_1^*(x,z)} =  i \partial_z \ln | \ell_1(z) |^2,
\end{equation}
where the symbol ($*$) denotes complex conjugation.

In one-dimensional systems, the action of the parity operator is defined as $\mathcal{P} f(x) = f(-x)$. In the two-dimensional case, there are two main definitions of the parity operator: reflection with respect to an axis and reflection with respect to the origin $\mathcal{P}_x f(x,z) = f( - x, z)$, $\mathcal{P}_2 f(x, z) = f(-x,-z)$. In this work, when the potential exhibits an explicit dependence on $z$, the latter definition will be applied, and we will simply use $\mathcal{P}$ instead of $\mathcal{P}_2$ for brevity. Furthermore, the action of the antilinear operator $\mathcal{T}$ is defined by $\mathcal{T} f(x,z) = f^*(x,z)$. 

To ensure that $V_1$ inherits the $\mathcal{PT}$ symmetry of $V_0$, we first consider the constraints imposed by the $\mathcal{P}_x \mathcal{T}$ condition:
\begin{equation} \label{TDDT-PxCondition}
    2\partial_x^2 \ln \frac{u_1(x, z)}{u_1^*(-x,z)} = i\partial_z \ln |\ell_1 (z)|^2.
\end{equation}
Similarly, requiring $V_1$ to inherit the $\mathcal{P}_2 \mathcal{T}$ symmetry of $V_0$ yields the condition:
\begin{equation} \label{TDDT-P2Condition}
    2\partial_x^2 \ln \frac{u_1(x,z)}{u_1^*(-x,-z)} = i\partial_z \ln \frac{\ell_1(z)}{\ell_1^*(-z)}.
\end{equation}

\subsection{Time-independent supersymmetric transformations} \label{TIDT}
     
The standard time-independent Darboux transformations can be recovered from the time-dependent framework by assuming that the initial potential is stationary, $V_0 = V_0(x)$. We employ separation of variables for the physical solutions, $\psi_0(x,z) = e^{-i E z} \phi_0(x)$, and for the transformation function, $u_1(x,z) = e^{-i \epsilon_1 z} v_1(x)$. Under these assumptions, the intertwining relationship in Eq.~\eqref{TDDT-S0S1IntertwineRel} reduces to 
\begin{equation} \label{TIDT-IntertwineRelation}
    H_1 A_1 = A_1 H_0, \qquad A_1(x) = \partial_x - \frac{\partial_x v_1(x)}{v_1(x)},
\end{equation}
where $H_i = -\partial_x^2 + V_i$ (for $i = 0, 1$) denote the stationary Hamiltonians. Furthermore, $\phi_0$ and $\phi_1$ are the specific eigenfunctions of $H_0$ and $H_1$, respectively. Here, $A_1$ is the time-independent intertwining operator. Consequently, the expression of the potential in Eq.~\eqref{TDDT-V1} simplifies to $V_1(x) = V_0(x) - 2\partial_x^2 \ln v_1(x)$, where the auxiliary function $v_1$ satisfies the stationary Schrödinger equation $ H_0 v_1 = \epsilon_1 v_1 $, with $\epsilon_1$ denoting the factorization energy. Finally, the solutions to the eigenvalue problem of the new Hamiltonian, $ H_1 \phi_1 = E \phi_1 $, are obtained by applying $A_1$ to the initial solutions, such that $\phi_1 = A_1 \phi_0$. As in the time-dependent case, it is required that $v_1$ be nodeless to prevent the potential $V_1$ from exhibiting singularities. 

The Hermiticity and $\mathcal{PT}$-symmetry conditions in Eqs.~\eqref{TDDT-HermiticCondition} and \eqref{TDDT-PxCondition} reduce to 
\begin{equation}
    \partial_x^2 \ln \frac{v_1(x)}{v_1^*(x)}  = 0, \qquad
    \partial_x^2 \ln \frac{v_1 (x)}{v_1^*(-x)}  = 0,
\end{equation}
respectively. 

\subsection{Time-dependent second-order supersymmetric transformations} \label{SOTDDT}

Given the Schrödinger operators $S_0$ and $S_1$ from Eq.~\eqref{TDDT-S0S1IntertwineRel}, which are intertwined by the first-order differential operator defined in Eq.~\eqref{TDDT-L1Operator}, we can iterate the time-dependent Darboux transformation. This allows us to intertwine $S_1$ with a new operator $S_2$, characterized by a potential $V_2(x,z)$ and solutions $\psi_2(x,z)$. The intertwining relation is given by
\begin{equation} \label{SOTDDT_S1S2Intertwine}
    S_2 L_2 = L_2 S_1, \qquad 
    L_2 = \ell_2 (z) \left (\partial_x - \frac{\partial_x w}{w} \right ),
\end{equation}
where $\ell_2(z)$ is an arbitrary non-vanishing function and $w(x,z)$ is a solution (not necessarily physical) to $S_1 w = 0$.  
As before, the solutions of $S_2 \psi_2 = 0$ are obtained via $\psi_2 = L_2 \psi_1$, provided that the intermediate functions $\ell_1$, $\ell_2$, $u_1$, and $w$ are regular. From Eqs.~\eqref{TDDT-S0S1IntertwineRel} and \eqref{SOTDDT_S1S2Intertwine}, the operators $S_0$ and $S_2$ are intertwined by the composite operator $L_{12} = L_2 L_1$. If we define $u_2(x,z)$ as the preimage of $w$ under $L_1$ such that $L_1 u_2 = w$, we can express $L_{12}$ and $V_2$ entirely in terms of the initial solutions $u_1$ and $u_2$. Note that $u_2$ is a solution to $S_0$, while its image $w$ satisfies $S_1 w = 0$. The resulting second-order operator is
\begin{equation} \label{SOTDDT_SODT}
    S_2 L_{12}  = L_{12} S_0, \qquad  
    L_{12}  = \frac{\ell_2 \ell_1}{W(u_1,u_2)} \left [ W(u_1,u_2) \partial_x^2 - W'(u_1,u_2) \partial_x + W(u_1',u_2') \right ],
\end{equation}
where the prime notation ($'$) explicitly denotes the partial derivative with respect to $x$, and $W(f, g) = f \partial_x g - (\partial_x f) g$ is the Wronskian of functions $f$ and $g$. The corresponding potential is given by
\begin{equation} \label{SOTDDT_V2}
    V_2 = V_0 - 2 \partial_x^2 \ln W(u_1,u_2) + i \partial_z \ln (\ell_1 \ell_2).
\end{equation}

A crucial consequence of choosing a preimage $u_2$ is that, even if the intermediate potential $V_1$ possesses singularities (due to zeros in $u_1$), the final potential $V_2$ can remain regular. Specifically, while $u_1$ may have zeros causing $V_1$ to be singular, $S_2$ will remain regular as long as the Wronskian $W(u_1,u_2)$ is nodeless. Under these conditions, the solutions of $S_2 \psi_2 = 0$ can be constructed directly from the solutions of $S_0 \psi_0 = 0$ via the transformation $\psi_2 = L_{12} \psi_0$.

The condition required to ensure that the final potential $V_2$ remains Hermitian (assuming a real $V_0$) is given by
\begin{equation} \label{SOTDDT_HermitianCondition}
    2 \partial_x^2 \ln \frac{W(u_1,u_2)}{W^{*}(u_1,u_2)} = i\partial_z \ln |\ell_1 \ell_2|^2.     
\end{equation}
Similarly, to ensure that $V_2$ exhibits strict $\mathcal{P}_2 \mathcal{T}$ symmetry (provided $V_0$ possesses the same symmetry), the required condition is
\begin{equation} \label{SOTDDT_P2TCondition}
    2 \partial_x^2 \ln \frac{W (u_1, u_2 )}{ \mathcal{P}_2 \mathcal{T} [ W(u_1, u_2) ] }  = i \partial_z \ln \frac{ \ell_1 (z)  \ell_2 (z) }{ \ell_1^* (-z) \ell_2^* (-z)}.
\end{equation}

\subsection{Time-independent second-order supersymmetric transformations} \label{SOTIDT}

The time-independent second-order Darboux transformation can be recovered from the general time-dependent framework in Sec.~\ref{SOTDDT} by assuming a stationary initial potential, $V_0 = V_0(x)$, and separable physical solutions, $\psi_0(x,z) = e^{-i E z} \phi_0(x)$. We similarly define separable transformation functions $u_j(x,z) = e^{-i\epsilon_j z} v_j(x)$ for $j=1,2$, where $v_j$ solves the eigenvalue equation of $H_0$ with factorization energies $\epsilon_j$. Under these assumptions, the arbitrary time-dependent functions $\ell_1(z)$ and $\ell_2(z)$ factor out, and the second-order operator $L_{12}$ reduces to the time-independent operator $A_{12}$, explicitly given by
\begin{equation} \label{SOTIDT_A12}
    A_{12} = \frac{1}{W(v_1,v_2)} \left [ W(v_1,v_2) \partial_x^2 - W'(v_1,v_2) \partial_x + W(v_1',v_2') \right ] .
\end{equation}
Consequently, the new Hamiltonian $H_2 = -\partial_x^2 + V_2$ is intertwined with the initial Hamiltonian $H_0$ via the relationship $   H_2 A_{12} = A_{12} H_0.$
The resulting potential $V_2$ is independent of the propagation variable $z$ and takes the form $V_2(x) = V_0(x) - 2\partial_x^2 \ln W(v_1,v_2).$

Finally, we consider the symmetry properties of the new potential. For $V_2$ to be Hermitian (assuming a real initial potential $V_0$), the transformation functions must satisfy the condition
\begin{equation} \label{SOTIDT_HermitianCondition}
    \partial_x^2 \ln \frac{W(v_1,v_2)}{W^*(v_1,v_2)} = 0.
\end{equation}
Alternatively, if $V_0$ is $\mathcal{P}_x \mathcal{T}$ symmetric, the condition for $V_2$ to inherit this symmetry is
\begin{equation} \label{SOTIDT_PxCondition}
	\partial_x^2 \ln \frac{W(v_1,v_2)}{\mathcal{P}_x \mathcal{T}[W^*(v_1, v_2)]} = 0.
\end{equation}

\section{Coupled waveguides: exact models and dynamics}

In this section, we generate exactly solvable models of coupled waveguides and study the dynamics of their guided modes, which will later serve as reference systems for assessing the accuracy of the tight-binding approach. To do so, we employ the techniques studied in \cite{contreras-astorgaPhotonicSystemsTwodimensional2019}. In the quantum-mechanical analogue, the coupled waveguides correspond to a double-well potential, it can be obtained after a second-order Darboux transformation described in Sec.~\ref{SOTDDT}, starting from the potential $V_0 = 0$. This corresponds to the free particle in quantum mechanics or, in the context of optics, a light beam propagating through a homogeneous medium.

The procedure requires selecting two transformation functions, $u_1$ and $u_2$, that satisfy the following criteria. First, they must be solutions to the free time-dependent Schrödinger equation, $(i\partial_z + \partial_x^2) u_i(x, z) = 0$. We construct these functions as a linear superposition of $N$ even and $M$ odd stationary solutions, defined as:
\begin{equation}
    u_i(x,z) = \sum_{n = 1}^N A_n \cosh (k_n x) \, e^{ik_n^2 z} + i \sum_{m = 1}^M B_m \sinh (k_m x) \, e^{i k_m^2 z},
\end{equation}
where $A_n$ and $B_m$ are real constants. The imaginary unit introduced in the odd terms ensures the $\mathcal{PT}$-symmetry of the transformation function. Second, the transformation functions $u_1$, $u_2$, along with the arbitrary functions $\ell_1$ and $\ell_2$ introduced in Sec.~\ref{SOTDDT}, must satisfy the Hermiticity or $\mathcal{PT}$-symmetry conditions outlined in Eqs.~\eqref{SOTDDT_HermitianCondition} and \eqref{SOTDDT_P2TCondition}. Finally, we require that the Wronskian $W(u_1,u_2)$ remains non-zero for all $x$ and $z$. This is a non-trivial regularity condition that will be verified for each specific example.

\subsection{Propagation-distance-independent Hermitian waveguides} \label{PDIHW}

To obtain two coupled waveguides, we select the transformation functions 
\begin{equation}
    v_1(x) = \cosh(k_1 x), \qquad 
    v_2(x) = i \sinh(k_2 x),
\end{equation}
where $k_1$ and $k_2$ are real constants. This choice naturally satisfies the condition given in Eq.~\eqref{SOTIDT_HermitianCondition}. Furthermore, $v_1$ and $v_2$ are eigenfunctions of the free Hamiltonian $H_0 = -\partial_x^2$ with factorization energies $\epsilon_1 = -k_1^2$ and $\epsilon_2 = -k_2^2$, respectively.

The potential derived using the explicit formula $V_S = -2\partial_x^2 \ln W(v_1,v_2)$ is 
\begin{equation} \label{PDIHWG_VHermitic}
    V_S(x) = \frac{(k_1^2 - k_2^2) \left [ k_2^2 (1 + \cosh 2k_1 x) - k_1^2 (1 - \cosh 2k_2 x) \right ]}{\left (k_1 \sinh k_1 x \sinh k_2 x - k_2 \cosh k_1 x \cosh k_2 x \right )^2}.
\end{equation}
This expression represents a symmetric double-well potential corresponding to two parallel waveguides, a system previously studied in Ref.~\cite{munoz-vegaExactlySolubleModel2014}. To ensure the regularity of the potential, the Wronskian $W(v_1,v_2)$ must be strictly free of zeros. Explicitly, the Wronskian is given by
\begin{equation}
    W(v_1,v_2) = k_2\cosh k_1 x \cosh k_2 x - k_1\sinh k_1 x \sinh k_2 x.
\end{equation}
Analyzing the zero-crossing condition $W(v_1,v_2) = 0$ leads to the relation $k_2 / k_1 = \tanh k_1 x \tanh k_2 x$. Since $|\tanh \xi| < 1$ for all real $\xi$, the product on the right-hand side is strictly bounded by unity, i.e., $|\tanh k_1 x \tanh k_2 x| < 1$. Consequently, it follows that $W(v_1,v_2) \neq 0$ provided that $|k_2| > |k_1|$.

We can construct two guided modes for this potential using the free-particle solutions given by 
\begin{equation} \label{PDIHWG_H0Sols}
	f_1(x) = \sinh(k_1 x), \qquad
	f_2(x) = \cosh(k_2 x),
\end{equation}
and applying the intertwining operator $A_{12}$ defined in Eq.~\eqref{SOTIDT_A12}. These solutions correspond to the fundamental mode $\psi_g$ and the first excited mode $\psi_e$:
\begin{equation}\label{PDIHWG_GroundState}
    \psi_g(x) = \frac{k_2 (k_2^2 - k_1^2) \cosh k_1 x}{W(v_1,v_2)}, \qquad
    \psi_e(x) = \frac{k_1 (k_2^2 - k_1^2) \sinh k_2 x}{W(v_1,v_2)}.
\end{equation}
Here, $\psi_g$ and $\psi_e$ are eigenstates satisfying the stationary equations $(H_2 - E_g) \psi_g = 0$ and $(H_2 - E_e) \psi_e = 0$, where $H_2 = -\partial_x^2 + V_S$, while $E_g = - k_2^2$ and $E_e = -k_1^2$ denote the corresponding energy eigenvalues. The ground state $\psi_g$ is symmetric ($\mathcal{P} \psi_g(x) = \psi_g (x)$), while the excited state $\psi_e$ is antisymmetric ($\mathcal{P} \psi_e(x) = - \psi_e(x)$), where the parity operator is defined as $\mathcal{P} f(x) \equiv f(-x)$. Consequently, we can construct states localized within the respective potential wells by taking linear combinations of $\psi_g$ and $\psi_e$:
\begin{equation} \label{PDIHWG_HermitianLR}
    \psi_{l,r} (x, z)= \frac{1}{\sqrt{2}} \left (e^{ik_2^2 z} \psi_g \pm e^{ik_1^2 z} \psi_e \right ),
\end{equation}
where the subscripts $l$ (with the $-$ sign) and $r$ (with the $+$ sign) denote the left-hand and right-hand sides of the symmetry axis ($x=0$), respectively. These new states, $\psi_l$ and $\psi_r$, are not eigenstates of the parity operator; instead, they are interchanged under parity transformations, such that $\mathcal{P} \psi_l = \psi_r$. Furthermore, since they are not eigenstates of the Hamiltonian, they are non-stationary (dynamically evolving) states. 

Examining Eq.~\eqref{PDIHWG_HermitianLR}, we observe that after a propagation distance equal to the beat length $z=T=2 \pi / (k_2^2 - k_1^2)$, the state returns to its initial configuration, acquiring the phase $\exp [(i2\pi k_1^2 ) / (k_2^2 - k_1^2)]$. Thus, the system exhibits a periodic spatial oscillation between $\psi_l$ and $\psi_r$, characterized by the spatial beat frequency $k_2^2 - k_1^2$. On the other hand, when $z=T/2$, the initial state $\psi_l$ (or $\psi_r$) transforms into $\psi_r$ (or $\psi_l$), up to an overall phase factor $\exp [(i\pi k_1^2) / (k_2^2 - k_1^2)]$.

\subsection{Propagation-distance-independent \texorpdfstring{$\mathcal{PT}$}{PT}-symmetric waveguides} \label{PDIPTW}

We construct two $\mathcal{PT}$-symmetric coupled waveguides by choosing the transformation functions 
\begin{equation}
    v_1(x) = \cosh k_1 x + i \alpha \sinh k_1 x, \qquad
    v_2(x) = i\sinh k_2 x, 
\end{equation}
where $k_1$ and $k_2$ are real constants. These transformation functions satisfy the condition given in Eq.~\eqref{SOTIDT_PxCondition}. Furthermore, $v_1$ and $v_2$ are solutions to the stationary free-particle Schrödinger equations $(H_0 - \epsilon_1)v_1 = 0$ and $(H_0 - \epsilon_2)v_2 = 0$, with factorization energies $\epsilon_1 = -k_1^2$ and $\epsilon_2 = -k_2^2$, respectively. 

The potential obtained using the formula $V_S = -2\partial_x^2 \ln W(v_1, v_2)$ is 
\begin{equation} \label{PDIPTW-V}
     V_S(x) = \frac{2(k_1^2 - k_2^2)}{W^2 (v_1,v_2)} \left [ k_2^2 (\cosh k_1 x + i\alpha \sinh k_1 x)^2 + k_1^2 (1 + \alpha^2) \sinh^2 k_2 x \right ].
\end{equation}
This potential corresponds to two $\mathcal{PT}$-symmetric waveguides exhibiting gain and loss, which are determined by its imaginary part, $\text{Im}[V_S(x)]$. To ensure the regularity of the potential, we must demonstrate that $W(v_1, v_2) \neq 0$. Computing the Wronskian of $v_1$ and $v_2$ yields
\begin{equation}
    k_2 \cosh k_2 x (\cosh k_1 x + i\alpha \sinh k_1 x)
   	- k_1 \sinh k_2 x (\sinh k_1 x + i\alpha \cosh k_1 x) = 0.
\end{equation}
We must identify the parameter values for which this expression vanishes. Separating the real and imaginary parts leads to the following conditions:
\begin{equation} 
    \frac{k_2}{k_1} - \tanh k_2 x \tanh k_1 x = 0, \qquad 
    \frac{k_2}{k_1} - \tanh k_2 x \coth k_1 x = 0.
\end{equation}
The real part condition can only be satisfied if $|k_2| < |k_1|$; it never vanishes when $|k_2| > |k_1|$. While the imaginary part has zeros for any choice of $k_1$ and $k_2$ (especially when $\alpha=0$), the non-vanishing of the real part is sufficient to guarantee the regularity of the potential. Consequently, the Wronskian satisfies $W(v_1, v_2) \neq 0$ provided that $|k_2| > |k_1|$ and $k_1 \neq 0$.

We can build two guided modes for this potential using the same free-particle solutions given in Eq.~\eqref{PDIHWG_H0Sols}. These solutions represent the $\mathcal{PT}$-symmetric counterparts of the fundamental mode $\psi_g$ and the first excited mode $\psi_e$:
\begin{subequations}\label{PDIPTW-Eigenstates}
\begin{align}
    \psi_{g}(x) &= N_g \frac{ k_2 (k_2^2 - k_1^2)}{W(v_1,v_2)} (\cosh k_1 x + i\alpha \sinh k_1 x), \label{PDIPTW-GroundState} \\
    \psi_{e}(x) &= N_e \frac{ k_1 (k_2^2 - k_1^2)}{W(v_1,v_2)} \sinh k_2 x, \label{PDIPTW-ExcitedState}
\end{align}
\end{subequations}
where $N_g$ and $N_e$ are normalization constants. These states satisfy the eigenvalue equations $(H_2 - E_g)\psi_g = 0$ and $(H_2 - E_e)\psi_e = 0$, where $H_2 = -\partial_x^2 + V_S$, with corresponding energy eigenvalues $E_g = -k_2^2$ and $E_e = -k_1^2$.

To obtain states localized in each potential well at $z=0$, we construct the linear combinations $\psi_l$ and $\psi_r$ by adding and subtracting $\psi_g$ and $\psi_e$:
\begin{equation} \label{PDIPTW-LR}
    \psi_{l,r}(x,z) = \frac{1}{\sqrt{2}} \left (e^{ik_2^2 z}\psi_g \pm e^{ik_1^2 z}\psi_e \right ), 
\end{equation}
where the subscripts $l$ and $r$ denote the left-hand and right-hand sides of the symmetry axis ($x=0$), respectively. These functions describe non-stationary states.

The funtions $\psi_l$ and $\psi_r$ in Eq.~\eqref{PDIPTW-LR} exhibit the same periodic exchange dynamics describes in Sec.~\ref{PDIHW}, with beat length $T=2 \pi/(k_2^2 - k_1^2)$ and spatial beat frequency $k_2^2 - k_1^2$. The non-Hermitian character of the present potential, however, introduces qualitatively distinct features in the energy flow, as we analyze later.

\subsection{Propagation-distance-dependent \texorpdfstring{$\mathcal{PT}$}{PT}-symmetric coupled waveguides} \label{PDDPTW}

To obtain two propagation-distance-dependent $\mathcal{PT}$-symmetric coupled waveguides, we start by choosing the transformation functions:
\begin{equation}
    u_1(x,z) = \cosh k_1 x \, e^{ik_1^2 z} + i\alpha \sinh k_3 x \, e^{ik_3^2 z}, \qquad
    u_2(x,z) = \sinh k_2 x \, e^{ik_2^2 z}, 
\end{equation}
where $k_1$, $k_2$, $k_3$, and $\alpha$ are real constants. Choosing $\ell_1=\ell_2=1$ satisfies the generalized $\mathcal{PT}$-symmetry conditions given in Eqs.~\eqref{TDDT-PxCondition} and \eqref{TDDT-P2Condition}. We observe that the system reduces to the Hermitian case discussed in Sec.~\ref{PDIHW} when $\alpha=0$, and reduces to the propagation-distance-independent $\mathcal{PT}$-symmetric case of Sec.~\ref{PDIPTW} by taking the limit $k_3\rightarrow k_1$. 

The potential is derived using the standard formula $V_S(x, z) = -2\partial_x^2 \ln W(u_1,u_2)$ defined in Eq.~\eqref{SOTDDT_V2}. This potential describes two coupled waveguides with balanced gain and loss, which are determined by $\mathrm{Im} [V_S]$. The structure exhibits periodicity in the $z$-coordinate with a spatial period $T_V=2\pi /(k_1^2 - k_3^2)$. For the potential to be physically valid, it must be free of singularities. This requires the Wronskian $W(u_1,u_2)$ to be nodeless for all $x$ and $z$. The explicit expression for the potential and the mathematical proof establishing the regularity condition ($|k_3|<|k_1|<|k_2|$), together with and additional bound on $\alpha$ given by $(1 - |k_1|/|k_2|)>|\alpha|(1 + |k_3|/|k_2|)$ are detailed in Appendix~\ref{App:PotentialReg}. We emphasize that this last bound is a conservative sufficient condition; in practice, the Wronskian remains nodeless and $V_2$ regular for a wider range of $\alpha$.

\begin{figure}[th!]
\centering
\includegraphics[width=.9\linewidth]{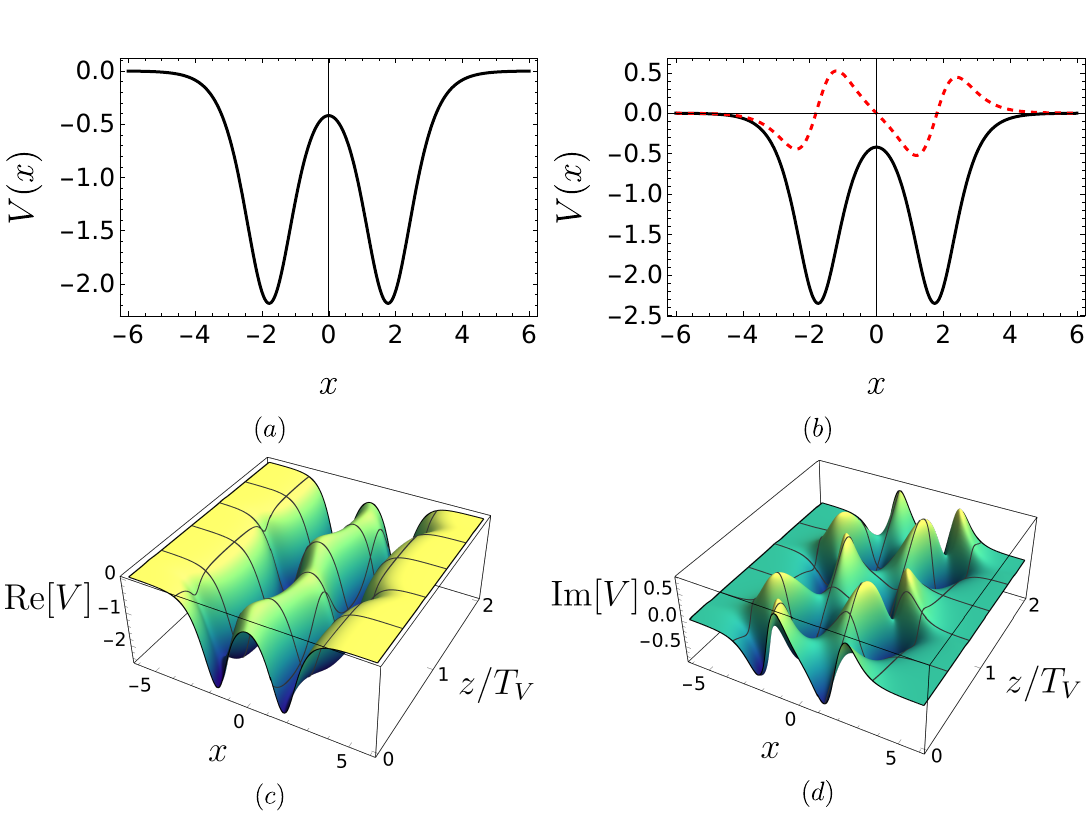}
\caption{Real and imaginary parts of the propagation-distance-dependent $\mathcal{PT}$-symmetric potential $V_S(x,z)$. Panels (a) and (b) depict transversal profiles associated with the limiting cases of the system: the strictly Hermitian potential ($\alpha = 0$) and the propagation-distance-independent $\mathcal{PT}$-symmetric potential ($k_3 \to k_1$) ($\mathrm{Re}[V]$ in black solid line and $\mathrm{Im}[V]$ in red dashed line), respectively. Panels (c) and (d) show the full three-dimensional $\mathrm{Re}[V]$ and $\mathrm{Im}[V]$ over two spatial periods $T_V$ along the propagation axis $z$, showing the periodic modulation of the coupled waveguides and their balanced gain and loss. Parameters: $k_1=1$, $k_2=1.1$, $k_3=0.95$, and $\alpha=0.5$. }
\label{fig:potentials}
\end{figure}

The dynamics of this potential are illustrated in Figure~\ref{fig:potentials}. Panel (a) shows the limiting Hermitian case $\alpha = 0$. In the limit where $k_3 \to k_1$, the $z$-dependence of the potential is completely suppressed, panel (b) shows the real (black solid line) and imaginary (red dashed line) parts of the potential in this case. Finally, the real and imaginary parts of the general potential $V_S(x,z)$ are plotted in panels (c) and (d), respectively. The complex potential exhibits periodic oscillations along the propagation axis $z$.    

We construct two guided modes for this system using the free-particle solutions
\begin{equation}
	f_1(x,z) = \sinh k_1 x \, e^{ik_1^2 z} + i\alpha \cosh k_3 x \, e^{ik_3^2 z}, \qquad
	f_2(x,z) = \cosh k_2 x \, e^{ik_2^2 z},
\end{equation}
and applying the second-order intertwining operator $L_{12}$ from Eq.~\eqref{SOTDDT_SODT}. This yields the non-stationary states:
\begin{equation} \label{PDDPTW-Psi_Def}
	\psi_1(x,z) = N_1 L_{12} f_1, \qquad 
	\psi_2(x,z) = N_2 L_{12} f_2,
\end{equation}
where $N_1$ and $N_2$ are normalization constants chosen such that the initial power of each mode is normalized to unity. These are computed by integrating the intensity at the input facet ($z=0$) using the generic formula:
\begin{equation} \label{PDDPTW-NConst}
    N_i = \left( \int_{-\infty}^{\infty} \big| \psi(x,0) \big|^2 dx \right)^{-1/2}.
\end{equation}
Due to their length, the explicit expressions of $\psi_1$ and $\psi_2$ are provided in Appendix~\ref{App:Modes}. These states are solutions to the paraxial wave equation, given by $\left(i\partial_z + \partial_x^2 - V_S\right)\psi = 0$.

Since the potential $V_S(x,z)$ is periodic in $z$, the solutions can be analyzed using Floquet theory \cite{dittrichQuantumTransportDissipation1998}. The states $\psi_1$ and $\psi_2$ correspond to two Floquet modes characterized by the quasi-energies $\epsilon_1 = -k_2^2$ and $\epsilon_2 = -k_1^2$, respectively. They can be expressed in the form $\psi_j(x,z) = e^{-i\epsilon_j z}\phi_j(x,z)$, where the functions $\phi_j(x,z)$ share the periodicity of the potential, i.e., $\phi_j(x,z+T_V) = \phi_j(x,z)$.

To analyze the coupling dynamics, we construct the localized states $\psi_l$ and $\psi_r$ (left and right of the symmetry axis at $z=0$) by superposing the Floquet modes:
\begin{equation} \label{PDDPTW-LR}
    \psi_{l,r}(x,z) = \frac{1}{N_{l,r}} ( \psi_1 \pm \psi_2 ),
\end{equation}
where $N_{l,r}$ are the specific normalization constants computed as in Eq.~\eqref{PDDPTW-NConst}.

While the potential has a fundamental period $T_V$, the localized superpositions $\psi_{l,r}$ generally exhibit a longer periodicity. After one potential period, the Floquet modes $\psi_1$ and $\psi_2$ acquire phase factors $e^{-i\Delta\theta_1}$ and $e^{-i\Delta\theta_2}$, with $\Delta\theta_1 = 2\pi k_2^2 /(k_1^2-k_3^2)$ and $\Delta\theta_2 = 2\pi k_1^2/(k_1^2-k_3^2)$. For the localized modes $\psi_{l,r}$ to return to their initial configuration, both phase factors must reduce to unity simultaneously. This requires the existence of integers $n$ and $m$ such that
\begin{equation}
    \frac{k_2^2}{k_1^2 - k_3^2} = \frac{n}{q}, \qquad \frac{k_1^2}{k_1^2 - k_3^2} = \frac{m}{q},
\end{equation}
for a common integer $q$. When this rationality condition holds, the full repetition period of the guided modes is given by $T_{\text{rep}} = \mathrm{lcm}(n,m) \, T_V$, where $\mathrm{lcm}(n,m)$ denotes the least common multiple of $n$ and $m$.

\section{Tight-binding models} \label{TBM}

Approximate methods have historically served as valuable tools for modeling complex physical systems. Among them is the tight-binding model, initially developed in solid-state physics to describe electron transport between atoms in a lattice \cite{kittelIntroductionSolidState2018}. This approach has since been adapted to optical systems, specifically for modeling light propagation in coupled waveguides \cite{amiriTightBindingAnalysisCoupled2006}. In this section, we briefly outline this theoretical framework. 

\subsection{Time-independent tight-binding model} \label{TBM-TI}

We begin with the time-independent Schrödinger equation:
\begin{equation} \label{TBM-TightBindingEq}
    H \psi(x) = \left (-\partial_x^2 + V(x) \right )\psi(x) = E \psi(x), 
\end{equation}
where $V(x)$ describes an array of potential wells, whose exact analytical solution may not be available. To find an approximate solution $\psi$ we start by expressing the potential as a superposition of individual wells:
\begin{equation}\label{TBM-VApprox}
    V_{TB}(x) = \sum_{j=1}^N V_0(x - x_j), 
\end{equation}
where $V_0$ is the potential associated with a single isolated well. 

We assume that the total wavefunction can be expressed as a linear combination of localized states. We introduce the localized basis functions $\phi_n(x) \equiv \phi_0(x - x_n)$, corresponding to the eigenfunctions of the isolated potential centered at $x_n$, which satisfy $H_0\phi_0 = \beta \phi_0$. The ansatz for the global solution is given by:
\begin{equation} \label{TBM-EigenApprox}
	\psi(x) = \sum_{j=1}^N c_j \phi_j(x) = \sum_{j=1}^N c_j \phi_0(x - x_j),  
\end{equation}
where $c_j$ are the expansion coefficients, $N$ is the total number of potential wells, and $x_j$ denotes the position of the $j$-th well. Substituting Eq.~\eqref{TBM-EigenApprox} into Eq.~\eqref{TBM-TightBindingEq}, multiplying from the left by the conjugate basis function $\phi_i^*(x)$, and integrating over all space yields a system of $N$ linear equations:
\begin{equation} \label{TBM-SystemEqs}
	\sum_{j=1}^N c_j H_{ij} = E \sum_{j=1}^N c_j S_{ij}, \qquad  i=1, \dots, N,  
\end{equation}
where the matrix elements are defined as:
\begin{equation} \label{TBM-MatrixElements}
    H_{ij} = \int_{-\infty}^{\infty} \phi_i^* H \phi_j \, dx, \qquad  
    S_{ij} = \int_{-\infty}^{\infty} \phi_i^* \phi_j \, dx.
\end{equation}
Here, $H_{ij}$ and $S_{ij}$ denote the Hamiltonian and overlap matrix elements, respectively. This constitutes a generalized eigenvalue problem, expressed in matrix notation as $\mathbf{H} \mathbf{c} = E \mathbf{S} \mathbf{c}$.

This generalized eigenvalue problem reduces to a standard one (where the overlap matrix becomes the identity, $\mathbf{S}=\mathbf{I}$) by orthogonalizing the basis functions $\phi_j$ via the Gram-Schmidt process:
\begin{equation}\label{TBM-GramSchmidt}
    \chi_n = \phi_n -\sum_{\mu=1}^{n-1} (\tilde{\phi}_{\mu}, \phi_n) \tilde{\phi}_{\mu}, \qquad n=1, \dots, N,  
\end{equation}
\begin{equation}\label{TBM-GramSchmidt2}
    \tilde{\phi}_n = \frac{\chi_n}{(\chi_n,\chi_n)^{1/2}}, \qquad (f,g) = \int_{-\infty}^{\infty} f^*(x) g(x) \, dx, 
\end{equation}
where $\tilde{\phi}_n$ represents the resulting orthonormal functions. Then, the differential eigenvalue equation \eqref{TBM-TightBindingEq} reduces to the standard numerical matrix equation $\mathbf{\tilde{H}}\mathbf{\tilde{c}} = E\mathbf{\tilde{c}}$ in the new basis, where the elements of the eigenvector $\mathbf{\tilde{c}}$ give the expansion coefficients of the wavefunction Eq.~\eqref{TBM-EigenApprox}, and the values $E$ are the approximate eigenvalues of Eq.~\eqref{TBM-TightBindingEq}. 

A fundamental assumption of the tight-binding framework is that the individual modes remain well-localized. Thus, the validity of the approximation rigorously depends on the spatial overlap integral between adjacent sites being small compared to unity. To quantify this for a pair of adjacent wells separated by a distance $2x_0$, we calculate the overlap parameter $\kappa$ between the single-well solutions from Eq.~\eqref{TBM-EigenApprox} centered at $-x_0$ and $x_0$:
\begin{equation} \label{TBHCW-kappa}
    \kappa = \int_{-\infty}^{\infty} \phi_0^*(x + x_0) \phi_0(x - x_0) \, dx.
\end{equation}

For non-Hermitian systems exhibiting $\mathcal{PT}$ symmetry, the standard inner product is no longer appropriate. Instead, the framework described above is consistently generalized by substituting the standard inner product with the $\mathcal{PT}$ inner product (or $\mathcal{P}$-pseudo norm), defined as:
\begin{equation} \label{TBM-PNorm}
    (f,g)_{\mathcal{PT}} = \int_{-\infty}^{\infty} f^*(x) \mathcal{P} g(x) \, dx .
\end{equation}
Consequently, the matrix elements in Eq.~\eqref{TBM-MatrixElements} and the Gram-Schmidt orthogonalization process in Eqs.~\eqref{TBM-GramSchmidt} and \eqref{TBM-GramSchmidt2} must be directly redefined with respect to this $\mathcal{PT}$ inner product.

\subsection{Time-dependent tight-binding model} \label{TDTB}

In Sec.~\ref{TBM-TI}, the solutions to the time-independent Schrödinger equation for Hermitian and $\mathcal{PT}$-symmetric periodic wells were approximated by expressing the total solution as a linear combination of the eigenstates of individual potential wells. It is possible to extend this approximation to the time-dependent Schrödinger equation by introducing a few modifications to the initial ansatz \cite{dellavalleSpectralTransportProperties2013}. Starting with the paraxial wave equation 
\begin{equation} \label{TDTB-CoupledEq}
    i\partial_z \psi(x, z) = H(x,z) \psi(x,z), \qquad H(x,z) = -\partial_x^2 + V(x,z),     
\end{equation}
where the potential $V(x,z)$ represents periodic symmetric potential wells in the transverse coordinate $x$ that change along the propagation distance $z$. We employ the same superposition ansatz as in Eq.~\eqref{TBM-EigenApprox}, 
\begin{equation}\label{TDTB-CoupledSol}
	\psi(x,z) = \sum_{j=1}^N c_j(z) \phi_0(x-x_j),
\end{equation}
with the crucial difference that the coefficients $c_j(z)$ are now $z$-dependent functions to be determined. The localized functions $\phi_0(x)$ remain solutions to the stationary Schrödinger equation $H_0\phi_0 = \beta \phi_0$, where $\beta$ is the corresponding eigenvalue.

Substituting Eq.~\eqref{TDTB-CoupledSol} into Eq.~\eqref{TDTB-CoupledEq}, we obtain 
\begin{equation}
	i \sum_{j=1}^N \phi_j(x) \partial_z c_j(z) = \sum_{j=1}^N H(x,z) \phi_j(x) c_j(z).
\end{equation}
Multiplying from the left by the complex conjugate $\phi_i^*(x)$ and integrating over the spatial coordinate $x$, we arrive at the system of coupled equations  
\begin{equation}
	i\sum_{j=1}^N S_{ij} \partial_z c_j(z) = \sum_{j=1}^N H_{ij}(z) c_j(z),
\end{equation}
with the matrix elements defined as
\begin{equation}
    H_{ij}(z) = \int_{-\infty}^{\infty} \phi_i^*(x) H(x,z) \phi_j(x) \, dx, \qquad S_{ij} = \int_{-\infty }^{\infty} \phi_i^*(x) \phi_j(x) \, dx.
\end{equation}
The elements $H_{ij}(z)$ and $S_{ij}$ represent $z$-dependent coupling coefficients and constant overlap integrals, respectively. This final matrix equation constitutes a system of coupled ordinary differential equations in $z$, which can be readily solved using standard numerical methods.  

\section{Tight-binding models of coupled waveguides}
\label{TBDW}

In this section, we apply the tight-binding method to the coupled waveguide systems introduced in Secs.~\ref{PDIHW}, \ref{PDIPTW}, and \ref{PDDPTW}. Our general strategy involves constructing single-well potentials and their corresponding eigenstates using first-order Darboux transformations. We employ free-particle solutions as the seed functions to generate both Hermitian and $\mathcal{PT}$-symmetric single wells. Subsequently, we approximate the coupled waveguide systems as a linear combination of these single-well potentials. For the propagation-distance system, this procedure allows us to reconstruct the oscillating guided modes by superposing the resulting Floquet states. Finally, we validate the approximation by comparing the results with the exact analytical solutions derived in the previous sections. In all examples, we will use the borderline overlap parameter $\kappa$, where the approximation starts showing discrepancies with the exact system.

\subsection{Tight-binding models for Hermitian coupled waveguides}
\label{TBHCW} 

\begin{figure}[th!]
\centering
\includegraphics[width = \linewidth]{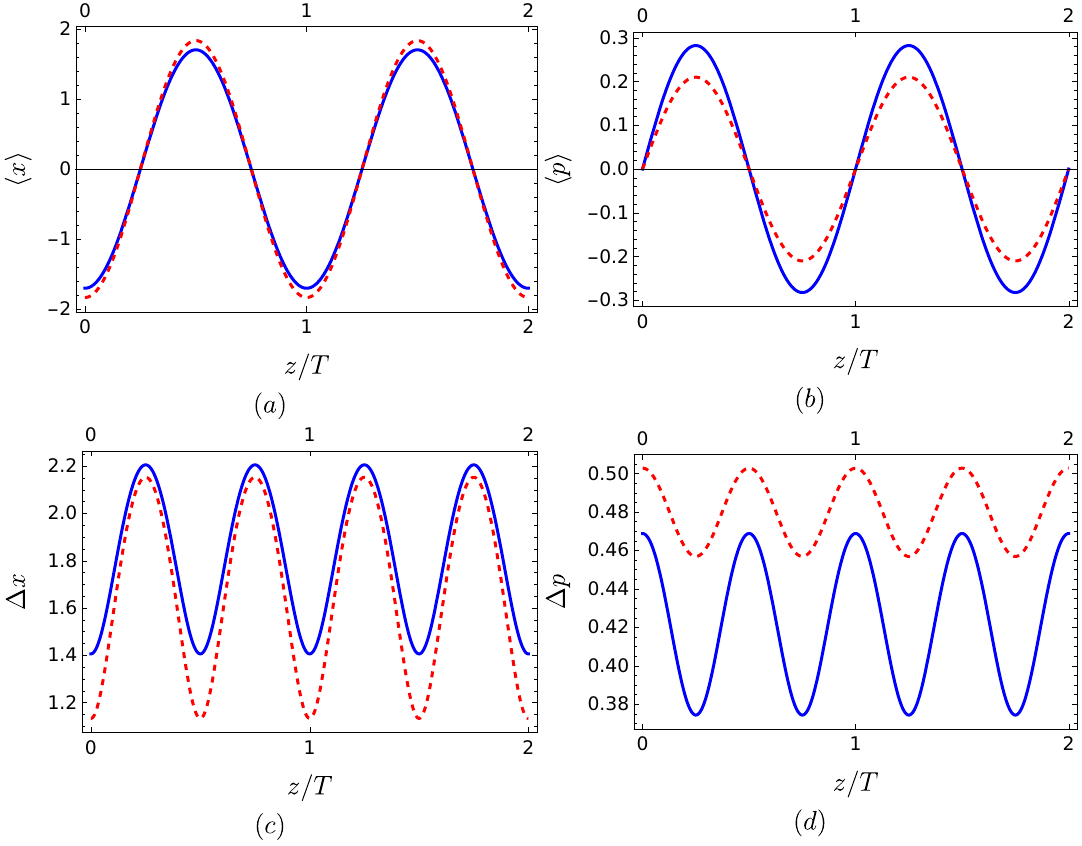}
\caption{Evolution of the observables for the guided mode $\psi_l$, comparing the exact solution (blue solid line) with the tight-binding model (red dashed line). Panels show: (a) position expectation value, (b) momentum expectation value, (c) position standard deviation, and (d) momentum standard deviation.}
\label{Chapter4_Expected_Hermitian}
\end{figure}

To model Hermitian coupled waveguides using the tight-binding framework, we first derive the properties of a single waveguide. We begin by selecting the transformation function $v(x)$, a solution to the free-particle equation given by $v(x)=\cosh(kx)$, where $k$ is a real constant, corresponding to a factorization energy $\epsilon=-k^2$. Since $\cosh(kx)$ is strictly positive for all real $x$, the resulting potential is regular everywhere and takes the form $V_0(x) = -2k^2 \sech^2(kx)$.

The fundamental mode of this potential, with energy $E=-k^2$, can be obtained by applying the operator $A_1$ defined in Eq.~\eqref{TIDT-IntertwineRelation} to the hyperbolic sine solution of the free particle, $f(x) = \sinh(kx)$. This yields:
\begin{equation} \label{TBHCW-Phi}
    \phi_0(x) = k\sech(kx).
\end{equation}

Using the superposition principle defined in Eqs.~\eqref{TBM-VApprox} and \eqref{TBM-EigenApprox}, we construct the approximate double-well potential $V_{TB}$ and the trial wavefunction $\psi$ as:
\begin{equation}\label{TBHCW-PotState}
	V_{TB}(x) = V_0(x - x_0) + V_0(x + x_0), \qquad \psi(x) = c_1 \phi_0(x - x_0) + c_2 \phi_0(x + x_0),
\end{equation}
where $x_0$ represents the displacement of each waveguide from the origin, and $c_1, c_2$ are the expansion coefficients to be calculated.

To determine the optimal tight-binding parameters $k$ and $x_0$, we calibrate the model against the exact system presented in Sec.~\ref{PDIHW}. Since the dynamical evolution and the tunneling period are governed by the energy splitting between modes, we employ a spectral matching method. Our objective is to minimize the deviation between the exact energy eigenvalues ($E_g=-k_2^2$ and $E_e=-k_1^2$) and the corresponding eigenvalues of the tight-binding Hamiltonian ($E_1$ and $E_2$), which depend on the parameters $k$ and $x_0$.

We define the cost function as the sum of the absolute energy differences. The optimization problem can be formulated as finding the parameters that satisfy:
\begin{equation} \label{TBHCW-Emin}
    \min_{k,x_0} \left \{|E_g - E_1(k,x_0)| + |E_e - E_2(k,x_0)| \right \},
\end{equation}
subject to the search intervals $x_0 \in (x_d - l, x_d + l)$ and $k \in (-k_0, k_0)$, where $x_d$ corresponds to the well separation of the exact potential and $l$ is a predefined search range. By numerically solving this minimization problem, we identify the optimal parameter set that best reproduces the physical spectrum.

Once the optimal parameters are determined, we construct the effective double-well potential $V_{TB}$ and its corresponding eigenstates. Using the eigenvectors obtained from the minimized Hamiltonian, we build the approximate fundamental and first excited modes. Finally, consistent with the procedure in Sec.~\ref{PDIHW}, the localized oscillatory modes are reconstructed by taking the symmetric and antisymmetric superpositions of these states.

In the following, we compare a specific example of two coupled waveguides modeled via the tight-binding model against the exact solution derived from a second-order Darboux transformation. We define the exact potential $V_S$ in Eq.~\eqref{PDIHWG_VHermitic} using the parameters $k_1=0.645$ and $k_2=0.865$. For the tight-binding model, the spectral matching method described by Eq.~\eqref{TBHCW-Emin} yields the optimized parameters $x_0=1.66214$ and $k=0.7454$. 

For our optimized parameters, we obtained the overlap parameter Eq.~\eqref{TBHCW-kappa}, $\kappa = 0.41$. While not infinitesimally small, this value appropriately characterizes a regime of non-negligible interaction, allowing us to test the robustness of the tight-binding model under significant waveguide coupling.

Figure~\ref{Chapter4_Expected_Hermitian} compares the dynamical evolution of the exact (blue solid line) and approximated (red dashed line) guided mode $\psi_l$ from Eq.~\eqref{PDIHWG_HermitianLR}. We analyze the expectation values for position $\braket{x}$ and transverse momentum $\braket{p}$, along with their respective standard deviations. Panel (a) illustrates the spatial oscillation of the energy distribution. The tight-binding approximation yields a slightly larger oscillation amplitude compared to the exact system, yet it accurately reproduces the spatial periodicity. Regarding the momentum evolution shown in panel (b), the approximation follows the oscillatory pattern of the exact mode, albeit with a slightly reduced amplitude. Crucially, it captures the phase behavior accurately: the momentum vanishes when the mode is maximally localized in one waveguide and reaches its extremum during the tunneling transit between wells. The localization properties are characterized by the standard deviations $\Delta x$ and $\Delta p$, shown in panels (c) and (d). The position uncertainty $\Delta x$ reaches its minimum when the mode is confined and its maximum when delocalized, with the approximation closely tracking the exact solution's range. Conversely, the momentum uncertainty $\Delta p$ exhibits the correct oscillatory behavior, although the approximation yields a systematic upward shift compared to the exact solution. Finally, since the Hamiltonian is Hermitian, quantities such as the energy expectation value $\braket{H}$ and the total power $P(z)$ are constants of motion. It is worth noting that this example operates in a regime with a non-negligible overlap integral ($\kappa=0.41$). While the tight-binding approximation converges to the exact solution as the waveguide separation increases, our analysis demonstrates that the model remains a reliable descriptor of the coupled-mode dynamics, accurately reflecting the evolution even when the interaction between waveguides is significant.

\subsection{Tight-binding models for propagation-distance-independent \texorpdfstring{$\mathcal{PT}$}{PT}-symmetric coupled waveguides}
\label{TBPTW}

To investigate time-independent $\mathcal{PT}$-symmetric coupled waveguides within the tight-binding framework, we construct the single-well potentials and their eigenstates using a first-order Darboux transformation. We begin by selecting the transformation function $v(x)$ as:
\begin{equation}
	v(x) = \cosh(kx) + i \tilde{\alpha} \sinh(kx),
\end{equation}
where $k$ is a real constant corresponding to the factorization energy $\epsilon=-k^2$. The resulting potential is given by:
\begin{equation} \label{TBM-V0PT}
	V_0(x) = -\frac{2k^2 (1 + \tilde{\alpha}^2)}{[\cosh(kx) + i \tilde{\alpha} \sinh(kx) ]^2}.
\end{equation}
This expression represents a complex optical waveguide, where the imaginary part, $\text{Im}[V_0]$, describes the transverse distribution of gain and loss. The potential is regular everywhere because the transformation function $v(x)$ is nodeless; specifically, its real part $\cosh(kx)$ is strictly positive for all real $x$. The fundamental mode of this potential is obtained by applying the operator $A_1$ defined in Eq.~\eqref{TIDT-IntertwineRelation} to the free-particle solution $f(x)=\sinh(kx)$:
\begin{equation} \label{TBM-Psi0}
	\phi_0(x) = C\frac{ k}{\cosh(kx) + i \tilde{\alpha} \sinh(kx)},
\end{equation}
where $C$ is a normalization constant evaluated with respect to the $\mathcal{PT}$ inner product defined in Eq.~\eqref{TBM-PNorm}.

Following the procedure established in the previous section, we construct the $\mathcal{PT}$-symmetric double-well potential and the corresponding trial wavefunction $\psi$ using Eqs.~\eqref{TBM-VApprox} and \eqref{TBM-EigenApprox}. The ansatz for the coupled system takes the form:
\begin{equation} \label{TBM-PsiPT}
	\psi(x) = c_1 \phi_0(x - x_0) + c_2 \phi_0(x + x_0).
\end{equation}
It is crucial to highlight a key feature of this construction: although the isolated single-well potential $V_0$ and its fundamental eigenstate $\phi_0$ do not inherently possess strict $\mathcal{PT}$ symmetry with respect to their own local centers, the global double-well potential $V_{TB}$---built from Eqs.~\eqref{TBM-VApprox} and \eqref{TBM-V0PT}---and the total superposition wavefunction $\psi$ successfully satisfy this macroscopic symmetry requirement with respect to the origin $x=0$.

To determine the optimal parameters $x_0$, $k$, and $\tilde{\alpha}$ for the tight-binding approximation, we employ the spectral matching strategy introduced in Sec.~\ref{TBHCW}. Our objective is to reproduce the exact energy spectrum ($E_g=-k_2^2$ and $E_e=-k_1^2$) by optimizing the tight-binding eigenvalues $\tilde{E}_1$ and $\tilde{E}_2$, which are functions of the variational parameters. The optimization problem is defined as minimizing the absolute deviation between the exact and approximate energy eigenvalues:
\begin{equation} \label{TBPTW-Emin}
    \min_{x_0,k,\tilde{\alpha}} \left \{ |E_g - \tilde{E}_1(x_0,k,\tilde{\alpha})| + |E_e - \tilde{E}_2(x_0,k,\tilde{\alpha})| \right \},
\end{equation}
subject to the search intervals $x_0 \in (x_d - l,x_d + l)$, $k \in (-k_0,k_0)$, and $\tilde{\alpha} \in (-\alpha_0,\alpha_0) $, where $x_d$ represents the exact well separation and $l, k_0, \alpha_0$ define the search bounds. Solving this minimization problem yields the optimal parameter set that best approximates the physical system.

Once the optimal parameters are identified, we reconstruct the fundamental and first excited states using the corresponding eigenvectors derived from Eq.~\eqref{TBM-SystemEqs}. Subsequently, the oscillating guided modes are generated following the procedure outlined in Sec.~\ref{PDIPTW}, by taking the appropriate linear combinations of the propagation-evolved eigenstates.

Following a parallel approach to the Hermitian case, we now present a comparative analysis for a $\mathcal{PT}$-symmetric coupled waveguide system. We compare the exact potential derived from a second-order Darboux transformation in Eq.~\eqref{PDIPTW-V} against the tight-binding approximation constructed using Eqs.~\eqref{TBM-VApprox} and \eqref{TBM-V0PT}. We begin by setting the exact system's parameters in Eq.~\eqref{PDIPTW-V} to $k_1=1.1$, $k_2=1.2$, and $\alpha=0.2$. For the tight-binding model, the effective parameters obtained via the spectral matching method are $x_0=1.65$, $k=1.14$, and $\tilde{\alpha}=0.21$. To ensure the applicability of the tight-binding framework, we evaluate the spatial overlap between the individual modes. Using Eq.~\eqref{TBHCW-kappa}, we calculate an overlap parameter of $\kappa=0.16$. Since this value is sufficiently small compared to unity ($\kappa \ll 1$), it confirms that the single-well states are well-localized, placing our chosen parameters firmly within the standard range of validity for the tight-binding model.

\begin{figure}[t!]
\centering
\includegraphics[width = \linewidth]{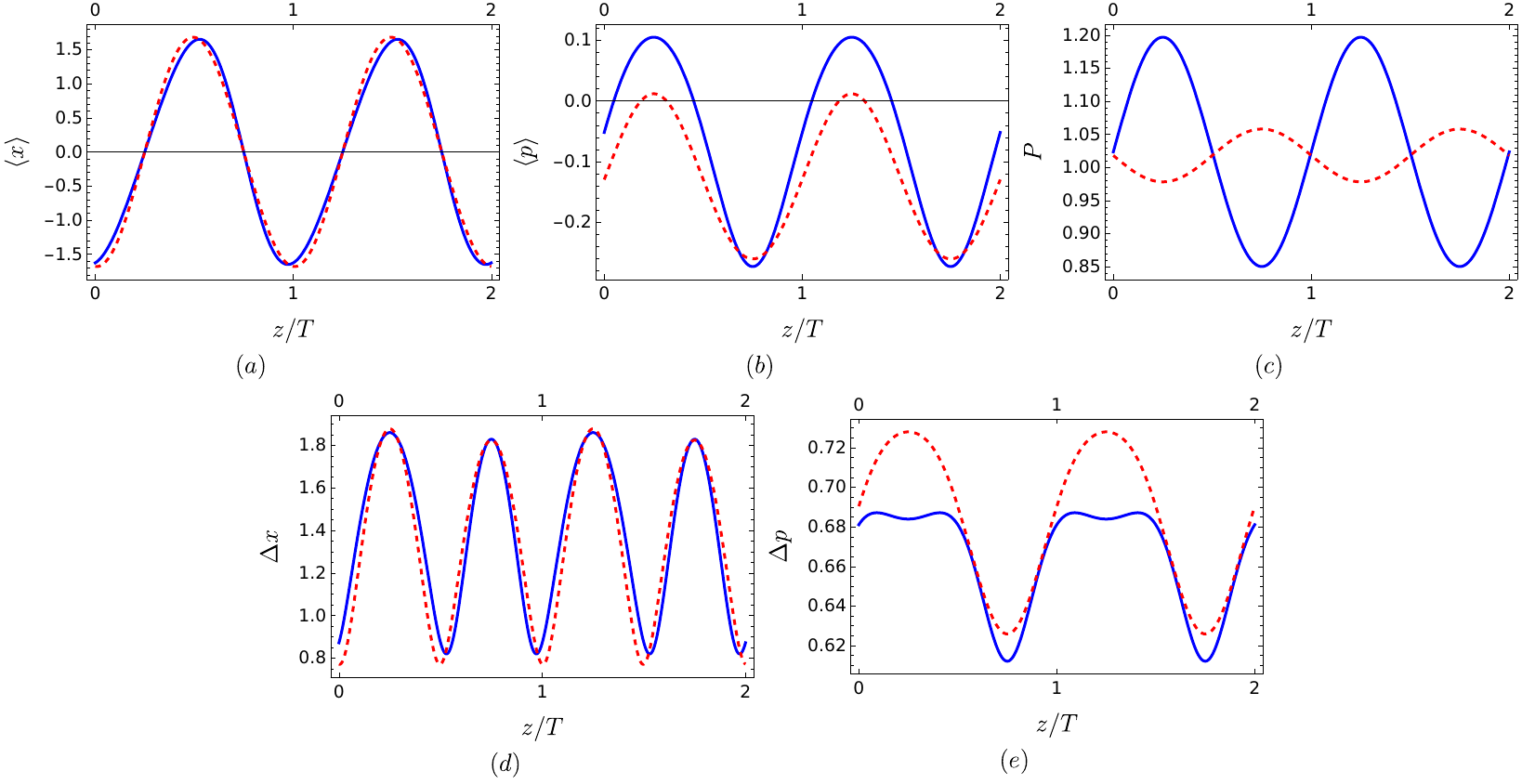}
\caption{Dynamical evolution of the guided mode $\psi_l$ for the exact (blue solid line) and approximated (red dashed line) models. Panels show the expectation values of (a) position, $\braket{x}$, (b) transverse momentum, $\braket{p}$, and (c) total power, $P$. Note the near-antiphase behavior of the power between the exact solution and the approximation. Panels (d) and (e) display the standard deviations of position, $\Delta x$, and momentum, $\Delta p$, respectively. These plots are generated using the localized states defined in Eq.~\eqref{PDIPTW-LR}.}
\label{TBPTW-ExpectedPT}
\end{figure}

Figure~\ref{TBPTW-ExpectedPT} compares the dynamical evolution of the exact (blue solid line) and approximated (red dashed line) guided mode $\psi_l$ from Eq.~\eqref{PDIPTW-LR}. We analyze the expectation values of position $\braket{x}$, transverse momentum $\braket{p}$, and total power $P$, along with the standard deviations of position and momentum. Panel (a) displays the expectation value of the position. The approximation exhibits a slightly larger oscillation amplitude compared to the exact solution, yet it effectively captures the spatial oscillation period of the energy distribution. Panel (b) shows the expectation value of the momentum. Unlike the Hermitian case, the momentum does not strictly vanish when the mode is maximally confined in one waveguide. This behavior is characteristic of non-Hermitian systems with complex potentials, reflecting continuous transverse energy fluxes. Despite a quantitative offset in the oscillation range, the tight-binding model qualitatively reproduces the system's dynamics: the momentum extrema still correspond to the regimes where the beam is delocalized and transferring between waveguides. Panel (c) depicts the total power evolution computed as
\begin{equation}
	P(z) = \int_{-\infty}^{\infty} |\psi(x, z)|^2 dx.
\end{equation}
 Both models predict an oscillatory power behavior, reflecting the periodic energy exchange with the active medium (gain and loss regions), a feature inherent to optical systems with complex refractive indices. It is highly notable that, while the approximation captures the periodicity, it exhibits a near-complete antiphase shift (approximately $\pi$) relative to the exact solution. Panel (d) presents the position standard deviation. The approximation yields values closely comparable to the exact solution. Similar to the Hermitian case, the spatial uncertainty is minimized during maximal confinement and maximized during the tunneling or delocalization phase. Finally, panel (e) shows the momentum standard deviation. The approximation accurately captures the magnitude and frequency of the fluctuations. This metric mirrors the behavior of the expected momentum, maintaining a non-zero spread even during maximal confinement, further highlighting the fundamental dynamic distinctions between Hermitian and $\mathcal{PT}$-symmetric modal propagation.

\begin{figure}[th!]
\centering
\includegraphics[width = \linewidth]{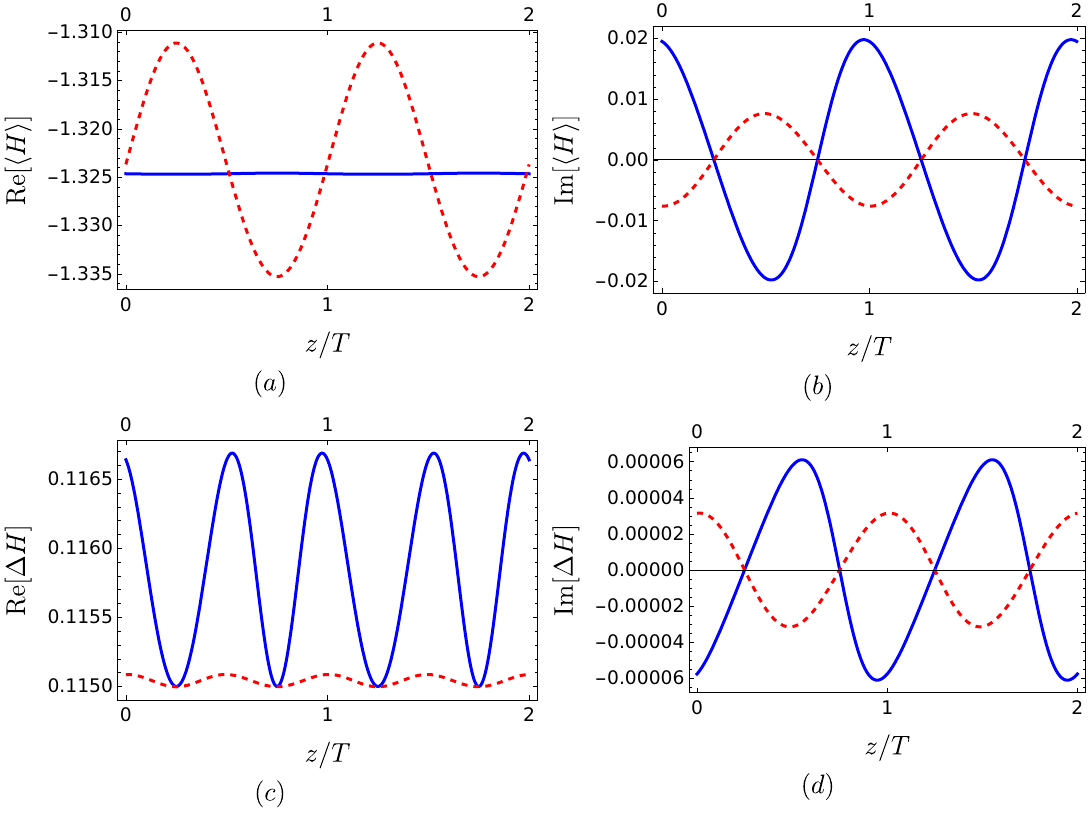}
\caption{Analysis of the expectation value of the Hamiltonian for the exact (blue solid line) and approximated (red dashed line) guided mode $\psi_l$. Panels show the (a) real and (b) imaginary components of the Hamiltonian expectation value, $\braket{H}$, calculated using the standard Hermitian inner product. Panels (c) and (d) display the real and imaginary components of the Hamiltonian uncertainty, $\Delta H$, respectively.}
\label{TBPTW-ExpectedEnergies}
\end{figure}

Figure~\ref{TBPTW-ExpectedEnergies} presents a comparison of the system's Hamiltonian expectation values and fluctuations for the exact (blue solid line) and approximated (red dashed line) models. Panels (a) and (b) display the real and imaginary parts of the expectation value of the Hamiltonian, $\braket{H}$, calculated under the standard (Dirac) Hermitian metric. The real part (a) exhibits oscillations where the approximation yields a noticeably higher amplitude compared to the exact solution. While the exact solution appears visually as a constant value on this scale, it actually oscillates with an amplitude approximately two orders of magnitude smaller than that of the tight-binding approximation. Similarly, the imaginary part (b) shows small-amplitude oscillations centered around zero, with the approximation exhibiting a pronounced phase shift and inverted amplitude behavior relative to the exact solution. The complex nature of this expectation value reflects the non-Hermitian character of the system: the real component relates to the effective propagation constant, while the non-zero imaginary component indicates the active presence of gain and loss mechanisms. These oscillations demonstrate that $\braket{H}$ is not conserved in the standard sense; rather, it fluctuates as the optical field periodically interacts with the amplifying and attenuating regions. Consistent with the total power analysis, the tight-binding approximation exhibits a significant phase discrepancy relative to the exact solution. 

In contrast, if we compute the expectation value using the $\mathcal{PT}$ inner product (or $\mathcal{P}$-pseudo norm), $\braket{H}_\mathcal{P}$, we find that this quantity corresponds to the arithmetic mean of the fundamental and first excited mode eigenvalues, $(E_g + E_e)/2$. This yields a strictly constant real value of $-1.135$ for both the exact system and the approximation. The application of the $\mathcal{P}$ operator renders the Hamiltonian pseudo-Hermitian ($H^{\dagger} = \mathcal{P} H \mathcal{P}^{-1}$), which ensures purely real eigenvalues and recovers a rigorous conservation law for the optical system operating within the unbroken $\mathcal{PT}$-symmetric phase. 

Finally, we analyze the fluctuations by examining the standard deviation (uncertainty) of the Hamiltonian, $\Delta H$, computed using the standard Hermitian norm. Panels (c) and (d) display the real and imaginary components of these fluctuations, respectively. The real part (c) of the exact solution exhibits significant periodic oscillations, reflecting dynamic changes in the spectral spread during propagation. In contrast, the tight-binding approximation fails to capture this dynamic behavior, yielding an almost flat response pinned near the lower bound of the exact solution's fluctuations. The imaginary part (d) shows oscillations with extremely small magnitudes for both models; however, the approximation displays a noticeable phase shift and an even smaller amplitude. This comprehensive comparison indicates that while the tight-binding model perfectly captures the average conserved pseudo-energy via the $\mathcal{P}$-norm, it struggles significantly to reproduce the fine, dynamical oscillatory structure of the Hamiltonian fluctuations present in the exact solution.

In summary, we have presented a comparative analysis of two $\mathcal{PT}$-symmetric coupled waveguides using the tight-binding approximation versus the exact analytical solution. We evaluated key observables, including position, transverse momentum, and the expectation value of the Hamiltonian, along with their respective standard deviations. Our results indicate that spatial observables, such as the position operator and beam width, are accurately reproduced by the approximation. However, non-Hermitian specific quantities, specifically the total beam power and the standard expectation value of the Hamiltonian, exhibit quantitative deviations, most notably a severe phase shift in their oscillatory behavior that the linear combination of isolated wells fails to capture.

It is important to contrast this system with the Hermitian example discussed in Sec.~\ref{TBHCW}. Unlike the previous case, the $\mathcal{PT}$-symmetric example operates in a regime where the two potential wells are well-defined and widely separated ($\kappa = 0.16$). This distinction serves a dual purpose: it validates the tight-binding method in its standard regime of applicability (weak coupling or separated wells) and simultaneously exposes its limitations when describing the intricate energy exchange dynamics induced by complex potentials.

A crucial factor governing the approximation's performance is the non-Hermiticity parameter $\tilde{\alpha}$. This parameter introduces a highly non-trivial structure to the imaginary part of the potential, which is challenging to accurately approximate using a simple linear superposition of single wells. We observed that the model performs adequately in the weak perturbative regime ($\tilde{\alpha} \ll 1$), yielding reasonable ranges for power and $\braket{H}$. However, for larger values, the discrepancies in reconstructing the exact imaginary potential profile lead to the observed dramatic phase shifts. We conclude that while the tight-binding model provides a highly reliable description of the macroscopic spatial dynamics in $\mathcal{PT}$-symmetric waveguides, extreme caution must be exercised when analyzing non-conserved energetic quantities, as the standard approximation does not fully resolve the complex phase evolution driven by the localized gain and loss mechanisms.

\subsection{Tight-binding models for propagation-distance-dependent \texorpdfstring{$\mathcal{PT}$}{PT}-symmetric coupled waveguides} \label{TBPDPTW}

In this section, we extend the tight-binding approximation to a system of two coupled waveguides characterized by a refractive index that depends on both the transverse coordinate $x$ and the propagation distance $z$. To implement the model, one could essentially employ either $z$-dependent or $z$-independent single-waveguide basis functions. Since both approaches yield comparable results, we select the propagation-distance-independent method for the sake of parametric simplicity.

We proceed by constructing the static basis functions following the same procedure as in the Hermitian case (see Sec.~\ref{TBHCW}). The total wavefunction is proposed as a time-dependent linear combination of these static modes:
\begin{equation} \label{TBPDPTW-PTStates}
	\psi(x,z) = c_1(z) \phi_0(x - x_0) + c_2(z) \phi_0(x + x_0),
\end{equation}
where $c_1(z)$ and $c_2(z)$ are the propagation-dependent coefficients to be determined. This ansatz represents a superposition of two single guided modes displaced by $x_0$ and $-x_0$ from the origin.

Unlike the stationary cases discussed previously, the $z$-dependence of the system implies that there is no static eigenvalue equation to serve as a reference for parameter calibration. Consequently, the utility of the potential comparison shifts from spectral matching to profile matching. We determine the optimal tight-binding parameters $x_0$ and $k$ by minimizing the maximum absolute deviation between the real parts of the exact potential $V_S$ (evaluated at the input facet $z=0$) and the superposition potential $V_{TB}$. This optimization problem is formulated as:
\begin{equation} \label{TBPDPTW-VDiff}
     \min_{x_0,k} \left \{ \max_{x \in (-d,0)} \left| \text{Re} [ V_S(x,0) - V_{TB}(x; x_0, k) ] \right| \right \},
\end{equation}
subject to the constraints $x_0 \in (-X,X)$ and $k \in (-K,K)$, where $d$, $X$, and $K$ are predefined search ranges. Since no static spectral data is available, this geometric fitting at the input plane ($z=0$) provides the necessary conditions to fix the basis functions.

Once the parameters $x_0$ and $k$ are determined via Eq.~\eqref{TBPDPTW-VDiff}, we derive the coupled-mode differential equations. Solving this system numerically allows us to determine the continuous evolution of the coefficients $c_{1,2}(z)$. By imposing periodic boundary conditions corresponding to the potential's longitudinal periodicity, we obtain the approximate Floquet states analogous to those in Eq.~\eqref{PDDPTW-Psi_Def}. Finally, the oscillating guided modes $\psi_l$ and $\psi_r$ are reconstructed by taking the symmetric and antisymmetric combinations of these approximate Floquet states, following Eq.~\eqref{PDDPTW-LR}.

We illustrate the time-dependent tight-binding approximation using the exact coupled waveguide system defined by the parameters $k_1=1$, $k_2=1.1$, $k_3=0.95$, and $\alpha=0.1$. The optimal tight-binding parameters, determined by minimizing the profile difference of the real potential, were found to be $x_0=1.77114$ and $k=1.045$. In this case, the overlap parameter is $\kappa = 0.18$, which falls comfortably within the model's range of validity.

\begin{figure}[t!]
\centering
\includegraphics[width = \linewidth]{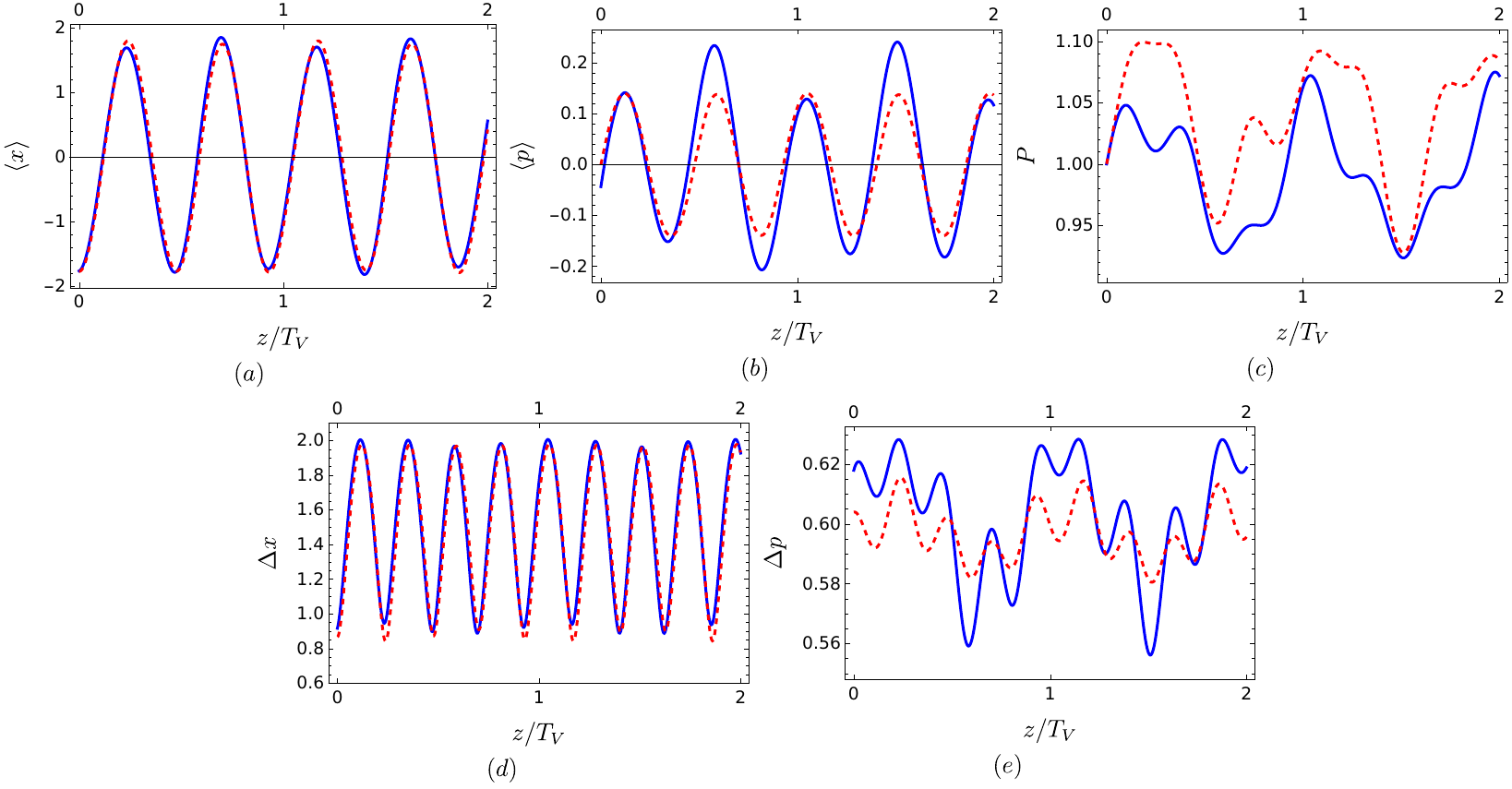}
\caption{Dynamical evolution of the guided mode $\psi_l$ for the exact (blue solid line) and approximated (red dashed line) solutions. Panels show the expectation values of (a) position, $\braket{x}$, and (b) transverse momentum, $\braket{p}$, along with (c) the total power, $P$. Panels (d) and (e) display the standard deviations of position, $\Delta x$, and momentum, $\Delta p$, respectively.}
\label{TBPDPTW-ExpectedPTTD}
\end{figure}

Figure~\ref{TBPDPTW-ExpectedPTTD} compares the dynamical evolution of the exact (blue solid line) and approximated (red dashed line) guided mode $\psi_l$ from Eq.~\eqref{PDDPTW-LR} by analyzing the expectation values of position $\langle x \rangle$ (a), transverse momentum $\langle p \rangle$ (b), total power $P$ (c), and the standard deviations of position $\Delta x$ (d) and momentum $\Delta p$ (e). The plots span two fundamental oscillation periods of the potential $V_S$ generated in Sec.~\ref{PDDPTW}.

Panel (a) illustrates the expectation value of the position. Here, the tight-binding approximation exhibits excellent agreement with the exact solution, accurately capturing the periodic spatial oscillation and confirming that the macroscopic transverse energy transfer between the waveguides is exceptionally well reproduced. Qualitatively, the expectation value of the momentum in panel (b) follows the behavior observed in the Hermitian case: extrema correspond to delocalization and zeros to confinement. However, the exact solution exhibits a more complex, multi-harmonic oscillatory structure compared to the smooth approximation. Panel (c) depicts the total power fluctuation. Although the behavior is inherently oscillatory, the full periodicity is not immediately apparent in the plot window. This occurs because the spatial beat period of the guided mode $\psi_l$ is significantly longer than the structural modulation period of the waveguides themselves (see Sec.~\ref{PDDPTW}). While the approximation captures the order of magnitude correctly, it reproduces the power dynamics only qualitatively, showing better agreement at integer multiples of the half-period. Finally, panels (d) and (e) show that the standard deviations exhibit regular oscillatory behavior. The spatial uncertainty is maximized when the guided mode is delocalized (distributed between waveguides) and minimized when the mode is highly localized, which is consistent with the fundamental confinement principles discussed in previous sections.

\begin{figure}[th!]
\centering
\includegraphics[width = \linewidth]{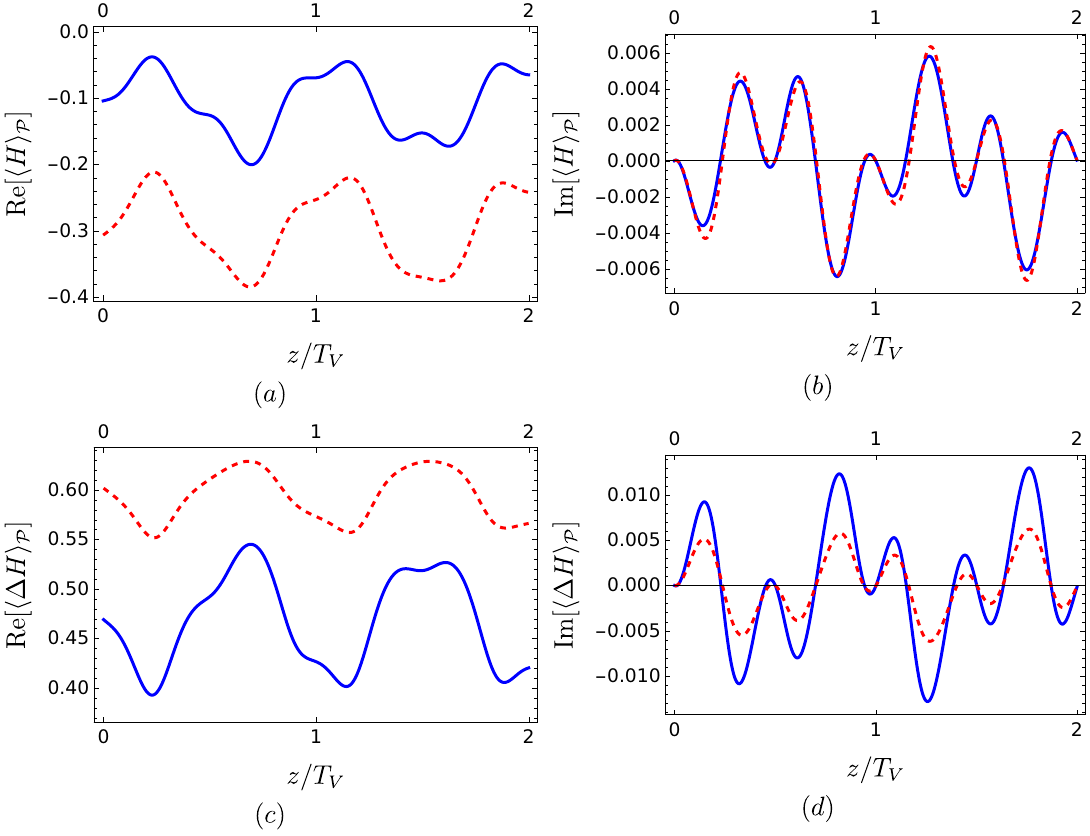}
\caption{Dynamical evolution of the system's energetics evaluated using the $\mathcal{P}$-inner product, comparing the exact (blue solid line) and approximated (red dashed line) solutions. Panels display the (a) real and (b) imaginary components of the expectation value of the Hamiltonian, $\braket{H}_\mathcal{P}$. Panels (c) and (d) show the real and imaginary components of the Hamiltonian standard deviation, $\Delta H_\mathcal{P}$, respectively.}
\label{TBPDPTW-ExpectedEnergies}
\end{figure}

Figure~\ref{TBPDPTW-ExpectedEnergies} presents a comparative analysis of the system's Hamiltonian expectation values and fluctuations for the exact (blue solid line) and approximated (red dashed line) solutions. We examine the real and imaginary components of the expectation value of the Hamiltonian, which correspond to the effective propagation constant and the net gain/loss dynamic contribution, respectively. For this analysis, we calculate the expectation values using the $\mathcal{PT}$-inner product (or $\mathcal{P}$-pseudo norm), normalized by the initial power, as these quantities provide a physically consistent description of the non-Hermitian system's integral behavior. Panels (a) and (b) display the real and imaginary parts of the expectation value of the Hamiltonian, $\langle H \rangle_\mathcal{P}$. The real part in the approximation (panel a) captures the general oscillatory dynamics but exhibits a consistent vertical offset relative to the exact solution. In contrast, the imaginary part (panel b) aligns remarkably well, confirming that the effective net gain/loss exchange is accurately represented by the tight-binding model. Regarding the spectral fluctuations, we present the standard deviation of the Hamiltonian, $\Delta H_\mathcal{P}$, in panels (c) and (d). The real part of this uncertainty (panel c) reveals a clear qualitative distinction and a consistent positive offset between the exact and approximated solutions, indicating that the spectral broadening is somewhat overestimated by the superposition ansatz. However, the imaginary part (panel d) shows notable qualitative agreement; it successfully reproduces the oscillatory frequency of the gain/loss uncertainty, although the approximation slightly underestimates the amplitude.

In summary, we examined the dynamical evolution of key observables---namely position, momentum, and the expectation value of the Hamiltonian---to evaluate the accuracy of the tight-binding approximation. Our analysis reveals that while the time-dependent tight-binding model successfully recreates the guided modes and their fundamental spatial oscillatory behavior, its direct quantitative prediction is limited for certain non-Hermitian quantities. Furthermore, regarding the system's spectral evolution, employing the $\mathcal{P}$-inner product is strictly essential for a physically consistent description. Under this metric, the approximation accurately reproduces the imaginary components associated with the net gain and loss dynamics; however, it captures the real components---representing the effective propagation phase and spectral broadening---only qualitatively, exhibiting a systematic vertical offset. Despite these quantitative discrepancies, the tight-binding model provides a highly valuable and computationally efficient conceptual framework for understanding the complex coupled dynamics of $z$-dependent $\mathcal{PT}$-symmetric optical structures.

\section{Conclusions} \label{Conclusion}

In this paper, we have presented a comprehensive comparative analysis between exact analytical solutions, derived via second-order Darboux transformations, and the tight-binding model for coupled optical waveguides with and without propagation-distance modulation. Our study encompassed three distinct configurations of coupled waveguides: Hermitian, propagation-distance-independent $\mathcal{PT}$-symmetric, and propagation-distance-dependent $\mathcal{PT}$-symmetric. These configurations allowed us to analyze both stationary and dynamically modulated systems within a unified theoretical framework.

Our results demonstrate that the tight-binding model remains a robust and efficient tool for describing light propagation in coupled waveguides, even in the presence of non-Hermitian gain and loss distributions. In terms of spatial dynamics, the approximation accurately reproduces the evolution of the guided modes, with the expected position and momentum closely tracking the exact analytical trajectories and capturing the essential localization phenomena. This agreement is particularly robust for spatial observables, even beyond the regime of extremely weak coupling. However, when analyzing energetic quantities, the model exhibits quantitative deviations. Notably, we identified a systematic phase shift in the oscillation of total power and energy expectation value in $\mathcal{PT}$-symmetric systems compared to the exact solution.

In conclusion, while the tight-binding model provides a physically intuitive and computationally inexpensive framework for $\mathcal{PT}$-symmetric optics, care must be taken when analyzing phase-sensitive quantities or systems with strong gain-loss contrast. The exact methods based on supersymmetric transformations serve as an essential benchmark, revealing the fine-structure dynamics that approximate models may overlook. These findings delineate the range of validity of discrete approximations for the design of non-Hermitian photonic devices and highlight the role of exact analytical methods as reliable benchmarks in complex modulated systems.

\section*{Acknowledgments}
ACA acknowledges Se\-cretar\'ia de Ciencia, Humanidades, Tecnolog\'ia e Innovaci\'on (SECIHTI - M\'exico) support under the grant FORDECYT-PRONACES/61533/2020.

\appendix
\section{Explicit expression of the potential and its regularity conditions} \label{App:PotentialReg}

The explicit form of the potential $V(x,z)$ discussed in Section \ref{PDDPTW} is given by:
\begin{equation}\label{App:PDDPTW-V}
    V(x,z) = \frac{h_3e^{i(k_1^2 - k_3^2)z} + \alpha^2 h_4 e^{i(k_3^2 - k_1^2)z} - i\alpha h_8}{h_1^2 e^{i(k_1^2 - k_3^2)z} - \alpha^2 h_2^2 e^{i(k_3^2 - k_1^2)z} + 2i\alpha h_1 h_2},
\end{equation}
where the auxiliary functions $h_1(x)$ through $h_8(x)$ are defined as:
\begin{align} 
	h_1(x) & = k_2 \cosh k_1 x \cosh k_2 x - k_1 \sinh k_1 x \sinh k_2 x, \\
    h_2(x) & = k_2 \cosh k_2 x \sinh k_3 x - k_3 \cosh k_3 x \sinh k_2 x, \\
    h_3(x) & = (k_1^2 - k_2^2 ) [2k_1^2 \sinh^2 k_2 x + k_2^2 (1 + \cosh 2 k_1 x)],\\
    h_4(x) & = 2(k_2^2 - k_3^2) (k_2^2 \sinh^2 k_3 x - k_3^2 \sinh^2 k_2 x),\\
    h_5(x) & = 2 k_1 k_3 (k_1^2 - 2 k_2^2 + k_3^2) \cosh k_3 x \sinh k_1 x \sinh^2 k_2 x,\\
    h_6(x) & = [ 4k_2^4 - 4k_1^2 k_3^2 \sinh^2 k_2 x + k_2^2 (k_1^2 + k_3^2) (\cosh 2 k_2 x - 3)] \cosh k_1 x \sinh k_3 x , \\
    h_7(x) & = k_2 (k_1^2 - k_3^2) (k_3 \cosh k_1 x \cosh k_3 x - k_1 \sinh k_1 x \sinh k_3 x) \sinh 2 k_2 x,\\
    h_8(x) & = h_5(x) + h_6(x) + h_7(x).
\end{align}

To ensure the potential is regular, the Wronskian $W(u_1, u_2)$ must be nodeless. Normalizing by the phase factor $e^{i(k_1^2 + k_2^2)z} $, the real and imaginary parts are:
\begin{align}
    \text{Re} \left [ \tilde{W} \right ] &
    = k_2 \cosh k_1 x \cosh k_2 x \left (1 - \frac{k_1}{k_2} \tanh k_1 x \tanh k_2 x \right ) \nonumber \\
    & + \alpha k_2 \sin [(k_3^2 - k_1^2)z] \cosh k_2 x \cosh k_3 x \left (\frac{k_3}{k_2} \tanh k_2 x - \tanh k_3 x \right ),
\end{align}
\begin{equation}
    \text{Im} \left [\tilde{W} \right ] = -\alpha k_2 \cos [(k_3^2 - k_1^2)z]
    \cosh k_2 x \cosh k_3 x \left (\frac{k_3}{k_2} \tanh k_2 x - \tanh k_3 x \right ). 
\end{equation}
The term in brackets in the imaginary part is monotonic and vanishes only at $x=0$. Thus, $\text{Im}[\tilde{W}] = 0$ only at $x=0$ or when the cosine vanishes. At these points, the real part becomes:
\begin{equation} \label{App:ZerosFunc}
    \frac{\cosh k_1 x}{\cosh k_3 x} \left (1 - \frac{k_1}{k_2} \tanh k_1 x \tanh k_2 x \right ) + \alpha (-1)^n \left ( \tanh k_3 x - \frac{k_3}{k_2} \tanh k_2 x \right ).
\end{equation}
Assuming the ordering $|k_3|<|k_1|<|k_2|$, the first term is bounded from below by $1-|k_1|/|k_2|$. The second term is bounded by $\pm \alpha(1 + |k_3|/|k_2|)$. Therefore, the Wronskian is guaranteed to be nodeless if:
\begin{equation}
    |k_3|<|k_1|<|k_2|, \quad k_2 \neq 0, \quad \text{and} \quad \left(1 - \frac{|k_1|}{|k_2|} \right) > |\alpha| \left(1 + \frac{|k_3|}{|k_2|} \right).
\end{equation}
We stress that this inequality is a sufficient condition, derived from worst-case bounds on the hyperbolic terms appearing in $\tilde{W}$. In practice, the potential $V(x,z)$ in Eq.~\eqref{App:PDDPTW-V} remains regular for a wider range of $\alpha$.

\section{Explicit expressions of the guided modes} \label{App:Modes}

The explicit expressions for the guided modes $\psi_1$ and $\psi_2$ derived in Section \ref{PDDPTW} are:
\begin{equation} \label{App-Psi1}
    \psi_1(x,z) = \frac{N_1 e^{2ik_2^2 z} k_2}{W(u_1,u_2)} \left [ e^{ik_1^2 z} (k_2^2 - k_1^2) \cosh k_1 x + i\alpha e^{ik_3^2 z} (k_2^2 - k_3^2) \sinh k_3 x \right ],
\end{equation}
\begin{equation} \label{App-Psi2}
    \psi_2(x,z) = \frac{N_2 e^{i(k_1^2 + k_2^2 + k_3^2) z}}{W(u_1,u_2)} \left [ e^{i(k_1^2 - k_3^2)z} k_1 (k_1^2 - k_2^2) \sinh k_2 x + \alpha^2  k_3 (k_3^2 - k_2^2) \sinh k_2 x - i\alpha K \right ],
\end{equation}
where the auxiliary function $K(x)$ is defined as:
\begin{equation}
    K(x) = k_2 (k_1^2 - k_3^2) \cosh k_2 x \cosh [(k_1 - k_3) x] + (k_1 + k_3)(k_1 k_3 - k_2^2) \sinh k_2 x \sinh [(k_1 - k_3)x].
\end{equation}

\bibliography{references}

\end{document}